\documentclass[sigconf]{acmart}

\AtBeginDocument{%
  \providecommand\BibTeX{{%
    \normalfont B\kern-0.5em{\scshape i\kern-0.25em b}\kern-0.8em\TeX}}}

\setcopyright{acmlicensed}
\copyrightyear{2018}
\acmYear{2018}
\acmDOI{XXXXXXX.XXXXXXX}

\acmConference[Conference acronym 'XX]{Make sure to enter the correct
  conference title from your rights confirmation emai}{June 03--05,
  2018}{Woodstock, NY}
\acmISBN{978-1-4503-XXXX-X/18/06}

\usepackage[noend]{algorithmic}
\usepackage{booktabs}
\usepackage{graphicx}
\usepackage{xspace}
\usepackage{subfig}
\usepackage{url}
\usepackage{listings}
\usepackage{algorithm}
\newcommand{\PaperAcronym}{LLMClean\xspace}
\newcommand{\quotes}[1]{``#1''}
\newcommand{\ra}[1]{\renewcommand{\arraystretch}{#1}} 
\renewcommand{\arraystretch}{1.2} 
\lstdefinestyle{promptstyle}{
  basicstyle=\small\ttfamily,
  columns=fullflexible,
  breaklines=true,
  backgroundcolor=\color{gray!10},
  frame=lr,
  framesep=8pt,
  framerule=0pt,
  xleftmargin=8pt,
  xrightmargin=8pt,
  captionpos=b,
  morekeywords={col_names, iot_names, yes, no},
  keywordstyle=\color{blue}\bfseries, 
  }
\lstdefinestyle{concept}{
    basicstyle=\ttfamily\small, 
    breakatwhitespace=false,
    breaklines=true,
    keepspaces=true,
}

\begin{document}

\title{\PaperAcronym: Context-Aware Tabular \\Data Cleaning via LLM-Generated OFDs}

\author{Fabian Biester}
\email{st108056@stud.uni-stuttgart.de}
\orcid{1234-5678-9012}
\affiliation{%
  \institution{University of Stuttgart}
  \streetaddress{P.O. Box 1212}
  \city{Stuttgart}
  \country{Germany}
}

\author{Mohamed Abdelaal}
\affiliation{%
  \institution{Software AG}
  \city{Darmstadt}
  \country{Germany}}
\email{Mohamed.Abdelaal@softwareag.com}

\author{Daniel Del Gaudio}
\affiliation{%
  \institution{University of Stuttgart}
  \city{Stuttgart}
  \country{Germany}
}

\renewcommand{\shortauthors}{Biester and Abdelaal, et al.}

\begin{abstract}
Machine learning's influence is expanding rapidly, now integral to decision-making processes from corporate strategy to the advancements in Industry 4.0. The efficacy of Artificial Intelligence broadly hinges on the caliber of data used during its training phase; optimal performance is tied to exceptional data quality. Data cleaning tools, particularly those that exploit functional dependencies within ontological frameworks or context models, are instrumental in augmenting data quality. Nevertheless, crafting these context models is a demanding task, both in terms of resources and expertise, often necessitating specialized knowledge from domain experts.

In light of these challenges, this paper introduces an innovative approach, called \PaperAcronym\footnote{Source code is available at \url{https://github.com/asdfthefourth/LLMClean}}, for the automated generation of context models, utilizing Large Language Models to analyze and understand various datasets. \PaperAcronym encompasses a sequence of actions, starting with  categorizing the dataset, extracting or mapping relevant models, and ultimately synthesizing the context model. To demonstrate its potential, we have developed and tested a prototype that applies our approach to three distinct datasets from the Internet of Things, healthcare, and Industry 4.0 sectors. The results of our evaluation indicate that our automated approach can achieve data cleaning efficacy comparable with that of context models crafted by human experts.
\end{abstract}


\keywords{Do, Not, Us, This, Code, Put, the, Correct, Terms, for,
  Your, Paper}

\received{20 February 2007}
\received[revised]{12 March 2009}
\received[accepted]{5 June 2009}

\maketitle

\section{Introduction}\label{sec:intro}
%
\paragraph*{Data Quality Problems} The data landscape is undergoing a massive expansion due to the proliferation of the Internet of Things (IoT), which is connecting an ever-growing network of data-gathering devices. This growth is supported by the declining costs of data acquisition technologies and the widespread availability of affordable internet, enabling devices to seamlessly integrate and communicate globally \cite{bansal2020survey}. Enterprises are increasingly harnessing this data to drive strategic business decisions and maintain a competitive edge, emphasizing the need for high-quality data. The escalating reliance on Machine Learning (ML) across diverse industries demands high-quality training data, where deficiencies in this data can lead to biased, inaccurate, or suboptimal ML outcomes~\cite{rein23}. Unfortunately, real-world data often contains various inaccuracies, e.g., duplications, null entries, anomalies, rule violations, and inconsistencies within or between data instances, all of which can substantially degrade the quality of the data.


\paragraph*{Context Awareness} To address data quality issues, a range of automated data cleaning tools have been developed, utilizing \textit{static} signals like business rules, data constraints, or metadata to identify and rectify errors in data \cite{holoclean17,raha19,holodetect19}. Despite their utility, these tools often lack incorporation of the context in which data is collected, a factor crucial for effectively cleaning data within ML workflows. Such context information provides insight into the data's meaning, relevance, and relationships, thereby ensuring that the cleaned data aligns accurately with the real-world phenomena it is intended to represent. To fill this gap, a suite of context-aware data cleaning tools, such as \cite{rtclean23, zheng2021discovery} have recently leveraged Ontological Functional Dependencies (OFDs) extracted from context models. In contrast to conventional functional dependencies, OFDs provide an advanced mechanism for capturing semantic relationships between attributes, which can significantly reduce the incidence of false positives while cleaning data (cf. Section~\ref{sec:context_model} for more details). 

\paragraph*{Challenges} OFD-based cleaning tools have demonstrated their efficacy in enhancing the precision of both error detection and correction. Nevertheless, the manual construction of context models for extracting OFDs is an inherently inefficient and impractical approach, particularly for real-time applications. This inefficiency stems from the need for extensive domain expertise to accurately interpret multifaceted and evolving data interrelationships, compounded by the overwhelming volume of data to be analyzed and the need for the models to rapidly adapt to environmental changes. Manual methods are further disadvantaged by their susceptibility to human error and limited scalability as system complexities increase. Moreover, ensuring consistency throughout the context model during updates presents an additional layer of difficulty. Therefore, the automation of this process is indispensable, not only to preserve the precision and trustworthiness of the context models but also to facilitate their scalability and flexibility amidst the swiftly changing data landscapes.

\paragraph*{Proposed Solution} In this paper, we introduce a novel method, designated as \PaperAcronym, which automatically generates context models from real-world data without requiring supplementary meta-information. \PaperAcronym leverages the powerful capabilities of Large Language Models (LLMs) to seamlessly adapt to dynamic data patterns. Specifically, \PaperAcronym includes several steps, including the classification of the dataset, the extraction or mapping of models, and the final generation of the context model. Thanks to the automatically generated OFDs, \PaperAcronym facilitates a robust data cleaning and analytical framework, addressing the challenges posed by the vast and evolving nature of real-world data, e.g., IoT datasets. Moreover, \PaperAcronym introduces a set of dependencies, namely Sensor Capability Dependencies and Device-Link Dependencies, pivotal for the precise detection of errors. Our evaluation shows that \PaperAcronym not only mirrors the data cleaning efficacy of manually curated context models but does so with enhanced efficiency and scalability. 

\paragraph*{Summary of Contributions} The paper provides the following contributions: (1) We introduce a novel three-stage architectural framework to identify erroneous instances in tabular data. This framework encompasses a comprehensive approach that combines the power of LLM models, context models, and data-cleaning tools. By leveraging this combined approach, our framework achieves significant improvements in both the effectiveness and efficiency of error detection compared to traditional tools, together with enhancing \PaperAcronym's ability to handle diverse and complex error patterns present in tabular data. (2) We present an innovative method that utilizes LLM models, such as Llama-2, GPT-3.5, and GPT-4, to autonomously generate context models directly from real-world data. (3) We propose an innovative prompt ensembling technique designed to enhance the stability of LLM models. (4) We develop an error detection tool that enforces a suite of OFD dependencies extracted from the automatically generated context models. (5) We conduct extensive experimental evaluation, comparing the performance of \PaperAcronym against a range of baseline methods using three real-world datasets from different domains, including IoT, Industry 4.0, and healthcare. To the best of our knowledge, \PaperAcronym is the first method that effectively leverages LLM models to enhance data cleaning tools through automatically generated context models. 

\paragraph*{Paper Structure} The remainder of this paper is structured as follows. Section~\ref{sec:overview} provides an overview of the \PaperAcronym method, outlining its key elements. Section~\ref{sec:context_model} introduces the proposed method for automating the context model generation using LLM models. Section~\ref{sec:prompt_ensembling} presents our prompt ensembling method to enhance the stability of LLM models. In Section~\ref{sec:cleaning}, we provide an overview of the error detection method developed to enforce the extracted OFD rules. Section~\ref{sec:evaluation} presents the experimental evaluation, including a discussion of results on standardized datasets and comparisons to baseline techniques. Section~\ref{sec:related_work} reviews related work on traditional data cleaning tools and distinguishes \PaperAcronym's novel formulation. Finally, Section~\ref{sec:conclusion} concludes and discusses potential directions for future extension.

\section{Overview}\label{sec:overview}
\begin{figure}
    \centering
    \includegraphics[width=1\linewidth]{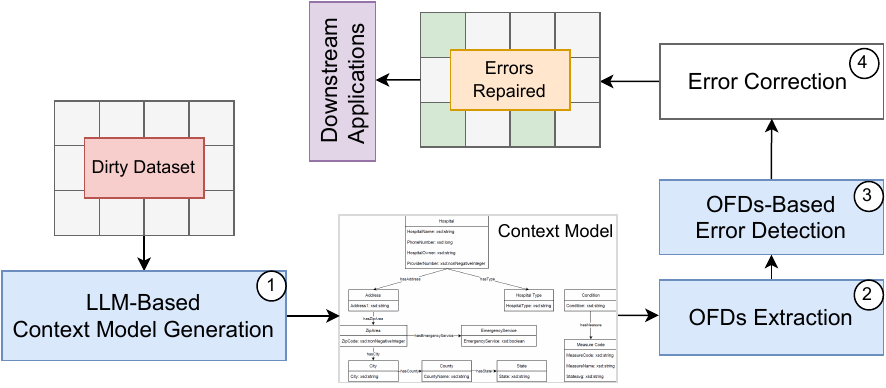}
    \caption{Architecture of \PaperAcronym}
    \label{fig:architecture}
\end{figure}

%
In this section, we introduce the architecture of \PaperAcronym together with relevant preliminaries. Figure~\ref{fig:architecture} shows that the input to the data cleaning pipeline is a dirty dataset, which may contain a heterogenous error profile, e.g., inaccuracies, inconsistencies, and missing entries. The data-cleaning process starts by generating a context model from this dirty dataset, which essentially maps out the critical relationships and attributes inherent within it. This model lays the foundation for the cleaning process ahead. Following the context model generation, \PaperAcronym identifies OFDs within the model—key indicators that signal potential data irregularities. \PaperAcronym leverages these OFDs to validate the input data. Data that pass this step are deemed valid, while the invalid data instances are flagged for further processing. Such information about the data being valid or not is later used as input to error correction tools, such as Baran\cite{raha19} and HoloClean\cite{holoclean17}, to generate repair candidates. 

By focusing on the erroneous instances identified by the OFDs, the error correction tools can systematically rectify errors, significantly boosting the dataset's overall quality. This seamless integration of automated tools and critical evaluations within the pipeline ensures the production of a dataset that is not only cleaner but also prepared for more reliable applications in various domains. Before delving into the automated generation of context models using LLMs, it is crucial to establish a clear understanding of the various types of OFD dependencies and how we categorize the input data as either \textit{IoT data} or \textit{non-IoT relational} data. 
%
\subsection{OFD Dependencies}
%
In general, OFDs represent a subset of Functional Dependencies (FDs) derived from an underlying Ontology, which provides the semantic framework necessary for establishing these dependencies. Ontologies serve as a formal representation of knowledge within a specific domain, providing a rich framework for defining the entities, relationships, and constraints that govern the data. This section introduces seven distinct types of OFDs that \PaperAcronym addresses, including denial dependency, matching dependency, device-link dependency, temporal dependency, location dependency, monitoring dependency, and capability dependency. Denial dependencies represent a broad category of integrity constraints that can express conditions disallowing certain data combinations, and this category includes the capability to represent constraints similar to functional dependencies (FDs) and conditional functional dependencies (CFDs)~\cite{holoclean17}. Formally, a denial dependency (DD) over a relation \( R \) is a constraint that denies the existence of certain tuples within an instance of \( R \). A denial dependency is expressed as a condition involving attributes of \( R \) that cannot hold simultaneously. For instance, a denial dependency \( D \) can be symbolically represented as \(\neg (X_1 \wedge X_2 \wedge \ldots \wedge X_n)\), where each \( X_i \) is a predicate over the attributes of \( R \). An instance \( I \) of \( R \) satisfies the denial dependency \( D \) if there are no tuples \( t1, t2, \ldots, tn \in I \) for which all predicates \( X_i \) are true simultaneously. If such a combination of tuples is found in \( I \), the involved tuples can be flagged as erroneous~\cite{rtclean23}. 

A Matching Dependency (MD) over a relation \( R \) is a constraint used to assess the correctness of data by evaluating the similarity between attribute values. An MD is denoted as \( A \overset{\sim}{\rightarrow} B \), where \( A \) and \( B \) are attributes within \( R \). The MD asserts that for every pair of tuples \( t1, t2 \in I \), a certain degree of similarity between \( t1[A] \) and \( t2[A] \) should imply a similarity between \( t1[B] \) and \( t2[B] \), according to a predefined similarity function. An instance \( I \) of \( R \) satisfies the matching dependency \( M \) if for every pair of tuples \( t1, t2 \in I \), the condition \( t1[A] \approx t2[A] \) implies that \( t1[B] \approx t2[B] \), where \( \approx \) denotes the similarity operator based on a specified similarity metric~\cite{rtclean23}. Aside from matching dependencies, a Device-Link Dependency \( L \) is defined by a mapping \( \Psi \) such that \( \Psi: X \rightarrow Y \), where \( X \) denotes the collection of sensors, and \( Y \) represents the ensemble of IoT devices. The dependency \( A \rightarrow B \), where \( A \in X \) and \( B \in Y \), is established if and only if sensor \( A \) is directly interfaced with device \( B \). This linkage enforces an exclusive read capability, meaning that data from sensor \( A \) can only be accessed through its linked device \( B \)~\cite{rtclean23}. Whereas, temporal dependencies \( T \) describe the sequencing of data transmission between devices \( A \) and \( B \). 

A temporal dependency \( A \rightarrow B \) signifies that device \( A \) precedes device \( B \) in time regarding data flow. Specifically, if a message \( m \) is timestamped at \( t_A \) when processed by \( A \), and subsequently at \( t_B \) by \( B \), then \( t_A < t_B \) must hold, reflecting the non-zero latency of transmission. A Location Dependency \( L \) is characterized by a mapping \( \Gamma: Device \rightarrow Location \), which associates sensing devices with their physical locations. Given a device \( A \) and location \( B \), such as a specific room, the dependency \( A \rightarrow B \) is established when \( A \) is positioned within \( B \), leading to \( \Gamma(A) = B \). Consequently, data collected by device \( A \) are indicative of the environmental conditions at location \( B \). Similarly, a Monitoring Dependency \( M \) describes the association between a device \( A \) and its monitoring entity \( B \). Within the IoT context, \( B \) tracks and records real-time health metrics of \( A \), such as CPU utilization and network connectivity. These measurements are captured and stored continuously as the system runs. A Capability Dependency \( C \) defines the relationship between a sensor \( A \) and its associated set of capabilities \( B \). Each capability is encapsulated as a metadata object linked to the sensor, specifying the type—like resolution or minimum measurable value—and the corresponding value for that type.

\subsection{Datasets Categorization}
\begin{figure}
    \centering
    \includegraphics[width=\linewidth]{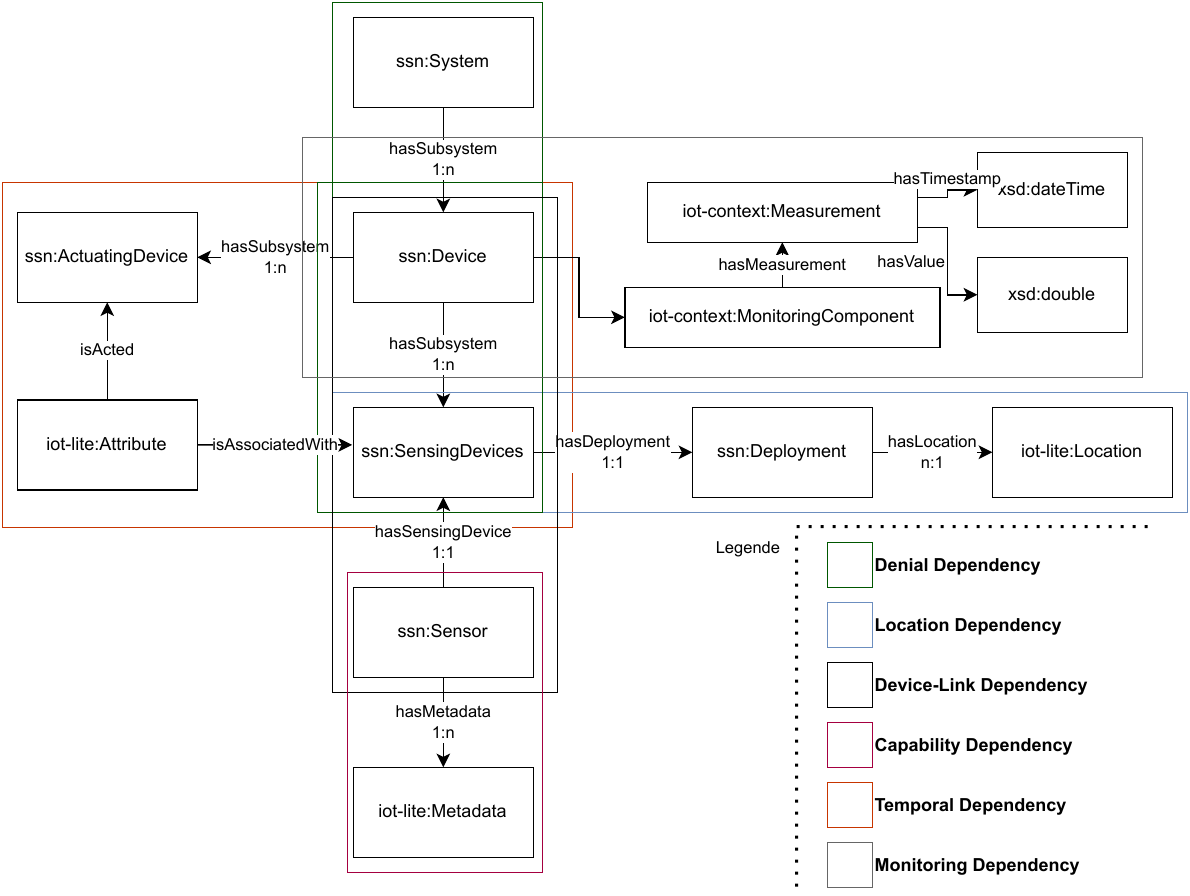}
    \caption{Metamodel of the context model for all IoT datasets}
    \label{fig:metamodel_iot}
\end{figure}

%
In \PaperAcronym, datasets serve as the primary input and are envisioned as collections of data points. These datasets should be structured in a single table format, complete with column headers. We categorize data into two principal classes, namely \textit{IoT datasets} and \textit{non-IoT relational datasets}, each with distinct requirements for the context model. Non-IoT datasets do not generally adhere to most OFDs as these dependencies are tailored to the architectural patterns of IoT sensors. Instead, only Matching and Denial dependencies are pertinent to non-IoT data. Conversely, IoT datasets comprise data from a network of interconnected sensors. For such datasets, dependencies unique to IoT, like Device-Link, Temporal, Locality, Monitoring, and Capability Dependencies, are relevant. These dependencies, which imply certain relationships within the data, inform the structure of the context model. Figure~\ref{fig:metamodel_iot} presents the meta-context model for IoT datasets, which consists of various concepts and their relationships. The colored boxes represent different dependencies and the associated concepts. 

The \lstinline[style=concept]{ssn:System} entity encapsulates the entire data-generating system, incorporating numerous \lstinline[style=concept]{ssn:device} entities that represent its subsystems. An \lstinline[style=concept]{ssn:device} may be a device such as a Raspberry Pi, a high-performance computer, or a standard PC, functioning in roles like actuator, sensor, or monitor. Monitoring components track system health metrics, such as CPU or network loads, encapsulated within \lstinline[style=concept]{iot-context:measurement} entities, each with a timestamp and value. Actuators integrate data across connected devices, while sensors are tied to \lstinline[style=concept]{ssn:device} entities and are deployed at specific locations. A single location can host multiple deployments, permitting several sensors within the same room but at distinct points. Sensors, which gather environmental data, are uniquely associated with a single sensing device and carry metadata detailing their operating range if such data is available. The Device-Link entity establishes the relationship between a sensor and its device through the sensing device. Similarly, the Capability Dependency links sensor metadata (like minimum and maximum operational values) to the sensor. Lastly, the Locality Dependency associates the sensing device with its deployment and physical location.
%
\section{Automated Context Modeling}\label{sec:context_model}
\begin{figure}
    \centering
    \includegraphics[width=1\linewidth]{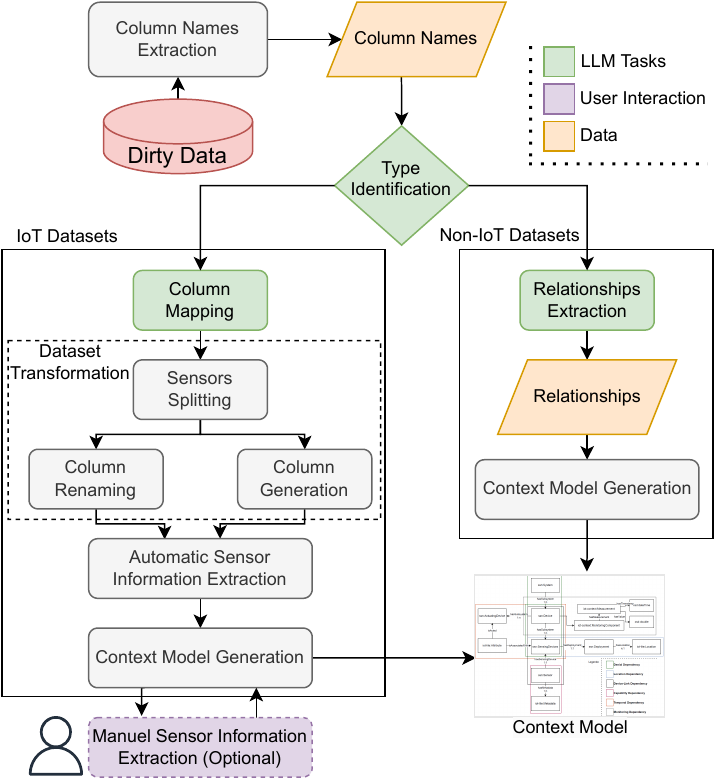}
    \caption{Automated generation of context model}
    \label{fig:context_model}
\end{figure}

%
In this section, we present a systematic approach for the automated generation of context models leveraging LLM models. Figure~\ref{fig:context_model} depicts the workflow that encompasses a sequence of steps tailored for both IoT and non-IoT relational datasets. The steps delineated in green signify the tasks where LLM models are employed to yield specific outcomes. The procedure commences with a dirty dataset, from which column names are extracted. These column names form the basis for the classification of the input dataset into respective types. Below, we elaborate on the steps specifically designed for each type of dataset.

\subsection{Handling IoT Datasets}

The workflow of IoT datasets initiates with column mapping. In this step, associations between the dataset's columns and the corresponding entities within the meta-context model (cf. Figure~\ref{fig:metamodel_iot}) are established. This mapping, critical for the subsequent generation of a context model, is executed via exploiting LLM models. The designated LLM model undertakes a systematic review of the predefined concepts, such as \lstinline[style=concept]{ssn:System}, \lstinline[style=concept]{ssn:Device}, \lstinline[style=concept]{ssn:SensingDevices}, \lstinline[style=concept]{ssn:Sensor}, \lstinline[style=concept]{iot-lite:Location}, \lstinline[style=concept]{iot-lite:Attribute}, \lstinline[style=concept]{ssn:ActuatingDevice}, \lstinline[style=concept]{iot-context:Measurement}, and \lstinline[style=concept]{iot-list:Metadata}, to determine their relevance to the columns at hand. In instances where a concept lacks a corresponding column, synthetic generation is employed to ensure completeness. Following successful column-to-concept correlation within the meta-model, the dataset undergoes a transformation phase to facilitate the creation of an actionable context model compatible with data-cleaning tools. This step is partitioned into three sub-steps, including sensor splitting, column renaming, and column generation.
\paragraph{Sensor Splitting} This sub-step is initiated upon the identification of multiple sensor readings within a single row in the input dataset during the column mapping step. In this sub-step, a composite dataset with multiple sensor readings per row is restructured into a singularized format. To illustrate, consider an initial dataset where each row is a tuple composed of temperature (T), CO2 concentration (C), location (L), and timestamp (t). The outcome of the sensor splitting sub-step is a dataset where each tuple's sensor readings are disaggregated into distinct rows. For instance, a row (T1, C1, L1, t1) in the original dataset is divided into two separate rows in the transformed dataset: one for temperature, (Temp, T1, L1, t1), and another for CO2 concentration, (CO2, C1, L1, t1). The location and timestamp for each sensor reading are replicated to maintain the integrity of the data, ensuring that each sensor value is contextualized by its original spatial and temporal information.
\paragraph{Column Generation} During the mapping step, there is a possibility that certain concepts may not be present in the input dataset or might not be recognized in the previous step. Such missing data may include parameters like the minimum and maximum sensor values, indicative of Capability Dependency, or data relating to the system's structural components, such as the device and sensor network details. The column generation phase is designed to resolve these gaps by introducing the requisite columns and populating them with synthetically derived values. However, it is pertinent to note that not all concepts or dependencies can be synthetically generated, leading to the potential exclusion of some concepts during this phase. Consider an example where the input dataset comprises only columns for \quotes{value}, \quotes{location}, and \quotes{timestamp}. Here, if \quotes{System}, \quotes{Device}, \quotes{SensingDevice}, and \quotes{Sensor} are requisite entities within the meta-context model, the absence of these columns necessitates their creation. Synthetic values are then assigned to these new columns to simulate system configuration. Moreover, to meet the requirements of Capability Dependency, additional columns like \quotes{MinValue} and \quotes{MaxValue} might be introduced, with default or synthetic ranges specified for sensor capacities.
\paragraph{Column Renaming} Upon successful assignment of columns to each concept, a validation is performed to ascertain whether the column titles align with the naming conventions requisite for the OFD generation phase. Discrepancies in column titles are rectified through a systematic renaming process, adhering to a pre-established schema. For instance, original column identifiers such as \quotes{Sensor\_name}, \quotes{temperature}, \quotes{place}, and \quotes{time} are systematically converted to \quotes{sensor}, \quotes{value}, \quotes{location}, and \quotes{timestamp}, respectively. This standardization of terminology facilitates seamless integration with the OFD extraction process, thereby sidestepping potential errors associated with inconsistent naming during subsequent data-cleaning operations.

After transforming the dataset, the automatic sensor information extraction step aims to identify the types of sensors employed and to establish the capability dependency within the context model by specifying the minimum and maximum operational values for each sensor. To accomplish this, queries are dispatched to resources such as LLM models, Wikipedia, and Wikidata. Additionally, \PaperAcronym provides an interface for end-users to optionally contribute sensor information directly to the context model, thereby enhancing its accuracy and comprehensiveness. This collaborative approach ensures that the context model remains robust and reflects the most current sensor capabilities. The workflow's final phase involves generating a concrete instance of the context model. To ensure data integrity, initial data-cleaning employs statistical methods to rectify potential errors, yielding sanitized entities. Subsequently, this refined dataset is structured into an RDF graph. Here, each dataset row manifests as a network of RDF triples, capturing the complex relationships and properties of sensor data in a semantic construct. This transformation process semantically augments the raw sensor data, enhancing its utility for applications dependent on semantically-enriched, high-quality datasets.

\subsection{Handling Non-IoT Datasets}
%
Constructing context models from non-IoT relational datasets necessitates a distinct approach from that employed for IoT data, utilizing only Matching and Denial dependencies. In this context, the workflow comprises three steps. First, all possible pairs of column names within the dataset are extracted. These pairs are subjected to analysis by LLM models to ascertain the presence of semantic relationships between the columns. Second, if a relationship between two columns is identified, the concept of the two columns is determined. Finally, \PaperAcronym assesses whether either column functions as an attribute of the other or stands as an independent concept. This critical evaluation serves to clarify the relationship between the columns, distinguishing whether they form part of a hierarchical arrangement or represent distinct concepts. Leveraging the data extracted from relationship extraction and concept mapping, the process stores column names as discrete concepts within the RDF graph. Relationships are then methodically established, linking less-encompassing concepts to more comprehensive ones. This approach facilitates the clear definition of hierarchical structures among the concepts, ensuring an organized and semantically meaningful RDF graph representation.
%
\section{Prompt Ensembling}\label{sec:prompt_ensembling}
%
In this section, we elaborate on the prompt engineering setup for querying LLM models. As discussed in Section~\ref{sec:context_model}, LLM models are exploited in multiple steps during the generation of context models due to their powerful reasoning capabilities. However, LLM models often tend to generate \quotes{hallucinated} or misleading results. To improve the outcomes of LLM models, \PaperAcronym leverages the concept of \textit{Prompt Ensembling}. The objective of prompt ensembling is to determine which combination of prompts yields the most accurate results when combined, as determined by a consensus method under various thresholds. To this end, the input dataset is  divided into two distinct subsets: a training set, employed to identify a collection of ensembles with high accuracy, as reflected by evaluation metrics like the F1 score, and their respective consensus thresholds; and a validation set, utilized to determine the optimal ensemble and its most effective threshold. 
\begin{figure}
    \centering
    \includegraphics[width=1\linewidth]{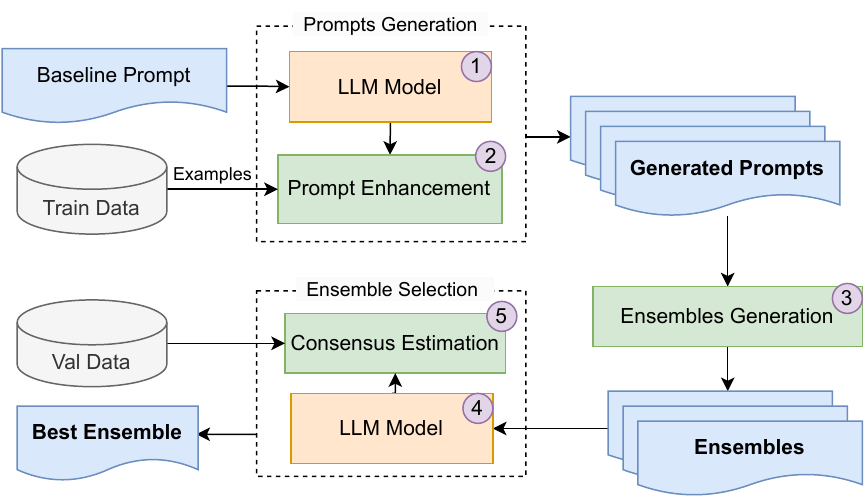}
    \caption{Prompt ensembling method}
    \label{fig:prompt_eng}
\end{figure}

Figure~\ref{fig:prompt_eng} shows the steps of our prompt ensembling approach. Initially, we craft a baseline prompt to guide an LLM model in producing a variety of prompts. The generated prompts are enhanced by adding carefully chosen examples from the training dataset, leveraging few-shot learning to better familiarize the LLM model with the expected response format. For instance, Listing~\ref{lst:type} illustrates a prompt for dataset type identification, comprising three elements: (1) few-shot examples, (2) a task description with input placeholders, and (3) a response format indicating whether a boolean or string answer is required. Algorithm~\ref{alg:ensemble} introduces the ensemble generation and the best ensemble selection procedures. At the outset, \PaperAcronym evaluates each prompt on the training and validation datasets and then explores all possible prompt combinations to form ensembles (lines~\ref{l:1},~\ref{l:2}). For each threshold value within a certain range, \PaperAcronym assesses every ensemble by aggregating the results of its constituent prompts. 
\begin{lstlisting}[style=promptstyle, caption={Prompt for data type identification}, label={lst:type}]
(1) Here are column names from an IoT dataset: {iot_names}.
(2) Do these names {col_names} suggest an IoT dataset?
(3) Answer with only yes or no.
\end{lstlisting}  
\begin{algorithm}
\caption{Find Best Ensemble Configuration}
\label{alg:ensemble}
\begin{algorithmic}[1]
\REQUIRE $train\_df$, $val\_df$, $prompts$, $tr\_range$
\ENSURE $best\_val\_config$
\STATE Compute prompt evaluations for $train\_df$ and $val\_df$ \label{l:1}
\STATE Generate $ensembles$ as all combinations of $prompts$ \label{l:2}
\FOR{each $threshold$ from 0 to $tr\_range$}
    \FOR{each $ensemble$ in $ensembles$}
        \STATE Collect $prompt\_results$ for prompts in $ensemble$ \label{l:3}
        \STATE Calculate $ensemble\_result$ using $find\_consens$ with $prompt\_results$ and $threshold$
        \STATE Find $f1$ score for $ensemble\_result$ on $train\_df$ \label{l:4}
        \IF{$f1 \geq best\_eval$} \label{l:5}
            \STATE Append $(threshold, ensemble)$ to $best\_configs$ \label{l:6}
        \ENDIF
    \ENDFOR
\ENDFOR
\FOR{each $config$ in $best\_configs$} \label{l:7}
    \STATE Collect $prompt\_results$ for prompts in $config[1]$ 
    \STATE Calculate $ensemble\_result$ using $find\_consens$ with $prompt\_results$ and $config[0]$
    \STATE Find $f1$ score for $ensemble\_result$ on $val\_df$
    \IF{$f1 \geq best\_eval$}
        \STATE Append $config$ to $best\_val\_config$ \label{l:8}
    \ENDIF
\ENDFOR
\RETURN $best\_val\_config$
\end{algorithmic}
\end{algorithm}
\begin{algorithm}
\caption{Finding Consensus Among Results}
\label{alg:consensus}
\begin{algorithmic}[1]
\REQUIRE $results$, $threshold$
\ENSURE $consens$
\STATE Initialize $result\_count$ using a counter overall $results$
\STATE Initialize $consens$ as an empty list
\FOR{each $obj, count$ in $result\_count$}
    \IF{$count \geq threshold$}
        \STATE Append $obj$ to $consens$
    \ENDIF
\ENDFOR
\RETURN $consens$
\end{algorithmic}
\end{algorithm}

\PaperAcronym then applies a function (introduced in Algorithm~\ref{alg:consensus}) to find a consensus among these results, considering the current threshold, and computes the F1 score for the ensemble on the training data (lines~\ref{l:3}-\ref{l:4}). Configurations that yield an F1 score at least as high as the best evaluation score recorded are retained as potential candidates (lines~\ref{l:5}-\ref{l:6}). After exhaustively evaluating all thresholds, the algorithm proceeds to test these top-performing configurations against the validation dataset. It appends those configurations that maintain or surpass the best evaluation score to a list of the best validation configurations (lines~\ref{l:7}-\ref{l:8}). The final output is this list, representing the ensemble configurations with the highest F1 scores on the validation data.

Algorithm~\ref{alg:consensus} provides a mechanism to achieve a consensus among results obtained from the ensemble. It requires a list of results and a threshold as inputs. The algorithm counts the occurrences of each result and compiles a consensus list. Through a voting mechanism, it identifies results that appear with a frequency that meets or exceeds the threshold, interpreting these as consensus results. These results are then compiled into a list that represents the collective decision of the ensemble. Together, these algorithms synergize to fine-tune the combination of prompts for better prediction accuracy and ensure that the results are robust by establishing a consensus based on a given threshold. 


\section{Data Cleaning with \PaperAcronym{}}\label{sec:cleaning}
%
In this section, we elaborate on how to leverage \PaperAcronym to detect errors in tabular data. As aforementioned, the generated context model is used to create a set of OFD rules. The core of our data-cleaning method lies in identifying two principal data quality issues: missing values and violations of functional dependencies within the data. The first aspect of the data-cleaning algorithm focuses on the ubiquitous issue of missing values, a commonly encountered rule in data management. Missing values are often represented by a variety of placeholders, e.g., \quotes{N/A}, \quotes{nan}, \quotes{none}, \quotes{null} or empty strings, that are not inherently recognized by standard data processing tools. 

\PaperAcronym{} addresses this challenge through exemplary rules such as \texttt{t1\&EQ(t1.System,\quotes{})}, which signifies a constraint where the \quotes{System} field in table \quotes{t1} should not contain empty strings. \PaperAcronym{} translates this rule into an actionable check by internally mapping these placeholders to \texttt{NaN}, thereby unifying all representations of missing data. This preliminary step is crucial for establishing a consistent ground truth from which missing data can be accurately identified. Subsequently, the algorithm iterates over the dataset to locate and record each occurrence of missing data, ensuring that no such instance escapes detection.

\sloppy Furthermore, \PaperAcronym{} excels at identifying more complex rule violations that involve functional dependencies between multiple fields across the dataset. Specifically, the value of one attribute (i.e., dependent) is supposed to be functionally determined by another attribute (i.e., determinant). Consider the rule \texttt{t1\&t2\&EQ(t1.SensingDevice,t2.SensingDevice)\&IQ(t1.De-}\\\texttt{vice,t2.Device)}. which asserts that for any pair of rows in tables \quotes{t1} and \quotes{t2}\footnote{In this context, \quotes{t1} and \quotes{t2} represent the same table.}, whenever the \quotes{SensingDevice} fields are equivalent, the \quotes{Device} fields must also be identical. Our algorithm enforces these constraints by first segmenting the dataset based on the determinant column. It then employs a statistical method to ascertain the modal value--the most frequently occurring dependent value within each group. This modal value is deemed the standard or legitimate value for a given determinant. Any deviation from this established norm is flagged as an anomaly.

By adopting this approach, our algorithm efficiently isolates and identifies instances where less common dependent values are present, which are likely indicative of data inconsistencies or errors. Through the intelligent application of these methods, we ensure that our algorithm effectively identifies violations of data integrity with precision. The error indices output by \PaperAcronym offers a clear and actionable guide for data practitioners to rectify the identified issues. Consequently, this approach significantly enhances the data-cleaning workflow, paving the way for more accurate and reliable data analyses.

%
\section{Performance Evaluation}\label{sec:evaluation}
%
In this section, we present an extensive evaluation of \PaperAcronym in different scenarios. Through a series of carefully designed experiments, we aim to address the following key questions: (1) How effective is the proposed prompt ensemble technique at achieving its intended outcomes? (2) To what extent does fine-tuning the Llama model enhance the efficacy of our prompt ensemble technique? and (4) How does the performance of \PaperAcronym, in terms of error detection and repair accuracy, compare to baseline methods?
 and  By addressing these questions, we shed light on the effectiveness and potential advantages of \PaperAcronym in the context of data cleaning. We first describe the setup of our evaluations, before discussing the results and the lessons learned throughout this study.


\subsection{Experimental Setup}
%
In this section, we introduce our experimental setup used while evaluating \PaperAcronym. We conducted our experiments on an Ubuntu 20.04 LTS machine, equipped with 256 cores @ 2.45 GHz, 1 TB RAM, and four Nvidia A100 GPUs with 40GB VRAM each. However, the minimum requirement is at least one GPU with 40GB of memory.
\paragraph{LLM Models} Several LLM models have been utilized in the evaluations. GPT4-turbo, the preview version, was selected for its enhanced performance and cost efficiency over GPT4. We also included GPT4, representing the series' fourth iteration, and GPT3.5, which was preferred in the development phase for its cost-effectiveness. These models provided a baseline for comparison. Additionally, we tested various Llama2 configurations—70b, 13b, 7b—to leverage its open-source accessibility and parameter-driven versatility for local execution tailored to specific computational needs. To enable faster inference time, quantization has been applied to the weights of Llama2-70b from 16 bits to 4 bits~\cite{lin2023awq}. To manage expenses, initial testing was conducted using GPT models on limited data samples, while comprehensive evaluations were predominantly performed using various versions of the Llama model.

\paragraph{Datasets} In the evaluations, we utilized three real-world datasets, namely \textit{IoT}, \textit{Hospital}, and \textit{CONTEXT} datasets, all provided as reference data with intentional errors to assess the data cleaning method's efficacy. Additionally, we utilize the \textit{LMKBC dataset} for evaluating the prompt ensembling method. In previous work \cite{rtclean23}, we gathered the IoT dataset, which includes context information. This dataset involved deploying three temperature sensors within a residential setting: two DS18B20 sensors in Room 1, one WSDCGQ11LM sensor in Room 2, and another WSDCGQ11LM sensor outside. The dataset consists of 1,000 entries, each organized into eight distinct columns: \quotes{System,} \quotes{Device,} \quotes{SensingDevice,} \quotes{Sensor,} \quotes{Name,} \quotes{Value,} \quotes{Timestamp,} and \quotes{Location.} To evaluate system robustness, we introduced 13\% numerical outliers and missing values, yielding 1041 erroneous instances. 

The CONTEXT dataset, sourced from a smart factory manufacturing electrical relays, encompasses data from five key stations: Inspection, Press, Robot, Transport Shuttle, and Storage \cite{kaupp2021context}. Each station features an array of sensors tracking diverse process parameters, resulting in a dataset with around 99,300 data entries, each organized into 22 columns. It also catalogs process errors, with a deliberate injection of a 0.05\% artificial error rate to simulate faults, resulting in 1219 erroneous instances. The Hospital dataset exemplifies a non-IoT, relational dataset depicting a U.S. hospital's operational data \cite{holoclean17}. It includes 1,000 rows across 19 columns and features a 2.6\% error rate through intentional data manipulation (resulting in 509 erroneous instances), serving as a robust dataset for evaluating \PaperAcronym. Table~\ref{tab:datasets} provides samples of the OFD rules extracted from the Hospital and IoT datasets.

\begin{table}
\centering
\ra{1.2}
\caption{Samples of the extracted OFD rules}
\label{tab:datasets}
\resizebox{\columnwidth}{!}{%
\begin{tabular}{@{}lll@{}}
\toprule
    & \textbf{IoT}  & \textbf{Hospital} \\ \midrule
OFD: Denial &
  \begin{tabular}[c]{@{}l@{}} Device $\to$ System,\\ SensingDevice $\to$ Device\end{tabular} &
  \begin{tabular}[c]{@{}l@{}}HospitalName $\to$ HospitalOwner,\\ ZipCode $\to$ City\end{tabular} \\
OFD: Matching &
  --- &
  \begin{tabular}[c]{@{}l@{}}ProviderNumber$_{75\%}$ $\to$ PhoneNumber$_{75\%}$,\\ Stateavg$_{75\%}$ $\to$ MeasureCode$_{75\%}$\end{tabular} \\
OFD: Device-Link      & ds18b20\_1 $\to$ device\_in\_1                                                                 & NA               \\
OFD: Capability       & \begin{tabular}[c]{@{}l@{}}ds18b20\_1 $\to$ MaxValue,\\ ds18b20\_1 $\to$ MinValue\end{tabular} & NA               \\
OFD: Locality         & \begin{tabular}[c]{@{}l@{}}ds18b20\_1 $\to$ Room1,\\ ds18b20\_2 $\to$ Room1\end{tabular}       & NA               \\
OFD: Temporal         & device\_in\_1 $\to$ device\_main                                                               & NA        \\ \bottomrule      
\end{tabular}%
}
\end{table}

In addition to the above datasets, we incorporated the LM-KBC dataset \cite{lmkbc23} as a benchmark for evaluating the efficacy of our proposed prompt ensembling algorithm. This particular dataset provides a rich variety of 21 distinct relations, each encompassing a different set of subject entities, coupled with a complete list of corresponding ground truth object entities for each subject-relation pair. The ML task associated with this dataset is to predict the object entities for each relation given a certain subject entity. The scope of the LM-KBC dataset is noteworthy, with relations spanning a multitude of domains such as chemistry, geography, and popular culture, among others. These relations are structured in a triple format: subject-predicate-object. To illustrate, within the chemistry domain, an example of such a relation is \quotes{CompoundHasParts,} connecting the subject entities \quotes{potassium, hydrogen, oxygen} with the object entity \quotes{Potassium Hydroxide.} In the context of our study, the training subset of the LM-KBC dataset is utilized for the initial few-shot training phase, while the validation subset plays a crucial role in determining the most effective configuration for our ensembling approach.

\paragraph{Evaluation Metrics} Our evaluations hinge on a carefully curated set of metrics to systematically assess data cleaning efficacy. For error detection, we leverage detection precision, recall, F1 score, and runtime to evaluate the effectiveness and efficiency. In this context, the
precision denotes the fraction of relevant instances, e.g., actual erroneous cells, among the detected instances. The detection recall is defined as the fraction of erroneous instances that
are detected. The detection F1 score denotes the harmonic mean of precision and recall. The runtime refers to the time taken to navigate through the dataset for error detection. It is pertinent to note that this figure excludes pre-processing activities for all tools, such as time spent generating OFDs for \PaperAcronym, setting up configurations for RAHA, and arranging FDs rules for HoloClean. This approach ensures a focused comparison of each tool's direct detection capabilities.

For the repairs, we differentiate between the numerical and the categorical attributes. For the former type, we employ the root mean square error (RMSE) as a distance measure between the repaired values and their ground truth.  For the latter data type, we employ precision, recall, and F1 measures. In this context, precision reveals the proportion of successful repairs against the aggregate of repair actions undertaken. In tandem, Recall captures the fraction of these correct repairs to the overall errors present in the data. 
Finally, the \quotes{Repairing F1 Score} emerges as a balanced metric, harmonizing precision with repairing recall, offering an analog to the traditional F1 Score but with a particular focus on the quality of the repair process. 
%
\subsection{Results}\label{sec:results}
%
In this section, we begin with evaluating the prompt ensembling algorithm, before presenting the evaluation results of \PaperAcronym in three scenarios, namely (S1) the \textit{Context Change} scenario, (S2) the \textit{Context Model} scenario, and (S3) the \textit{Sensor Capabilities} scenario.

\subsubsection{Prompt Ensembling}
\paragraph{Few-Shot Learning} In this experiment, we explore the effects of varying the number of examples utilized within the prompt on the overall performance. We conducted a series of experiments, each involving a random selection of few-shot examples from the training dataset for each relation. A careful selection process has been employed to ensure that the chosen few-shot examples spanned a comprehensive range, including those with the largest and smallest answer sets, as well as considering the possibility of empty sets where permissible. These examples have been then incorporated into either the best prompt (BP) or the best ensemble (BE) for evaluation.

\begin{figure}
    \centering
    \includegraphics[width=\columnwidth]{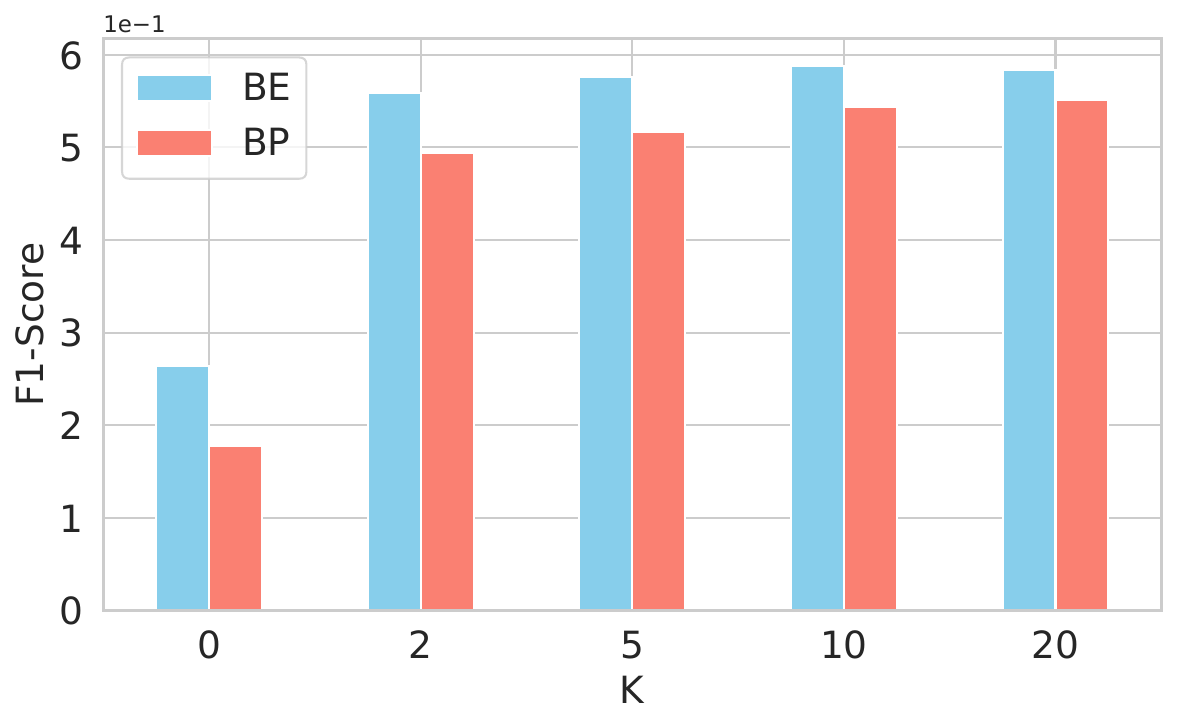}
    \caption{Impact of few-shot learning, where BE denotes the best ensemble and BP denotes the best prompt}
    \label{fig:few_shot}
\end{figure}

Figure~\ref{fig:few_shot} illustrates the outcomes of these experiments, specifically highlighting the variations in prediction accuracy as gauged by the F1-score with increasing example counts. The figure depicts a substantial enhancement in performance with the addition of examples to the prompts; the F1 score approximately doubles with the inclusion of merely two examples. Subsequent increments in the number of examples lead to more modest improvements in the F1-score. Furthermore, the figure highlights a consistently higher F1 score achieved through utilizing the best ensemble, surpassing the best prompt's performance by an average of 11.2\%. These findings not only demonstrate the value of increasing the number of few-shot examples for improving model accuracy but also underscore the superiority of using the best ensemble in leveraging these examples to achieve optimal results.

\begin{figure}
    \centering
    \includegraphics[width=1\columnwidth]{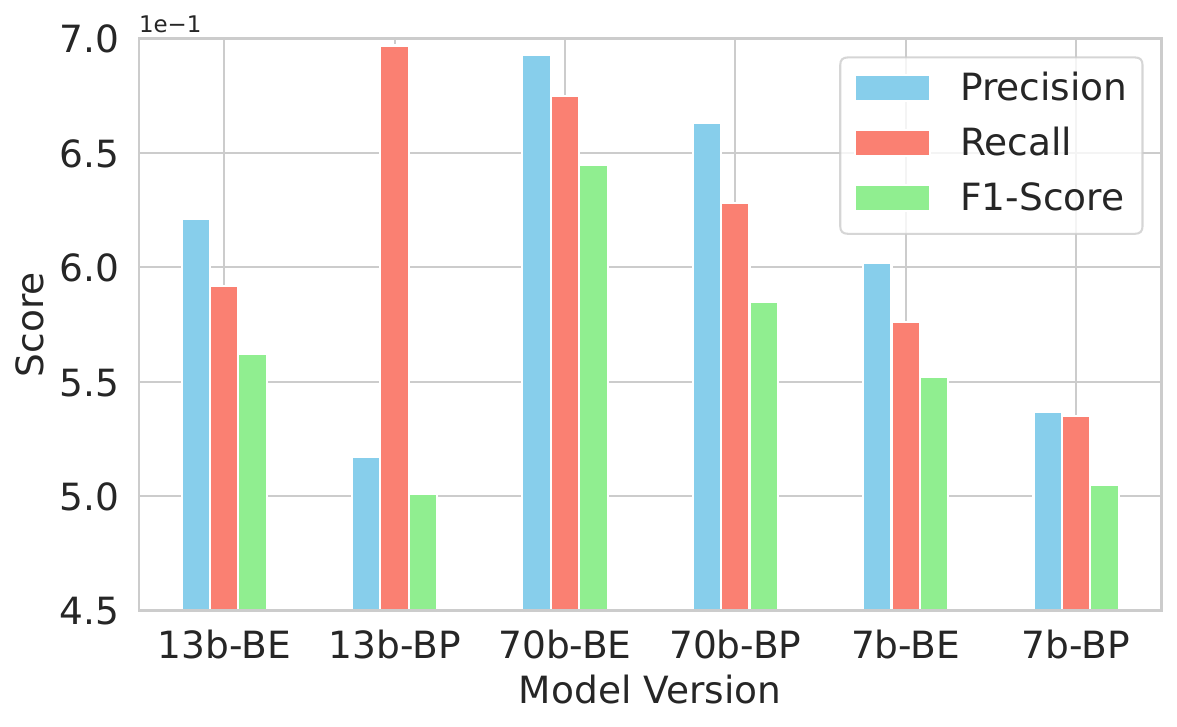}
    \caption{Comparisons of the Llama2 model with different parameter sizes, each with the best prompt or the best
ensemble}
    \label{fig:llama}
\end{figure}

\paragraph{Model Selection} In this set of experiments, we explored the performance of the Llama2 model across various configurations, examining editions with 7 billion, 13 billion, and 70 billion parameters. In addition to these, we compared the outcomes of a non-fine-tuned model against those of a version that had undergone fine-tuning specifically for chat completion tasks. As depicted in Figure~\ref{fig:llama}, our comparison focused on the differential impact of model size and the scenarios of using the best prompt and the best ensemble. The results from this figure indicate a clear trend: as the number of parameters in the Llama2 model increases, so does its performance. Notably, the 70 billion parameter variant, when combined with the best ensemble method, yields the highest F1 score. Another interesting insight from the figure is the fact that the 7 billion parameter model, when utilized in conjunction with the best ensemble, outperforms the 13 billion parameter model that employs only the best prompt. This observation suggests that the integration of the best ensemble technique can significantly elevate a model's effectiveness, to the extent that a less complex model can surpass a more complex one that does not utilize this optimization. These findings serve to emphasize the multifaceted nature of model performance, which hinges not only on the sheer scale of parameters but also critically on the strategic enhancements applied to the model's deployment.

\begin{figure}
    \centering
    \includegraphics[width=1\columnwidth]{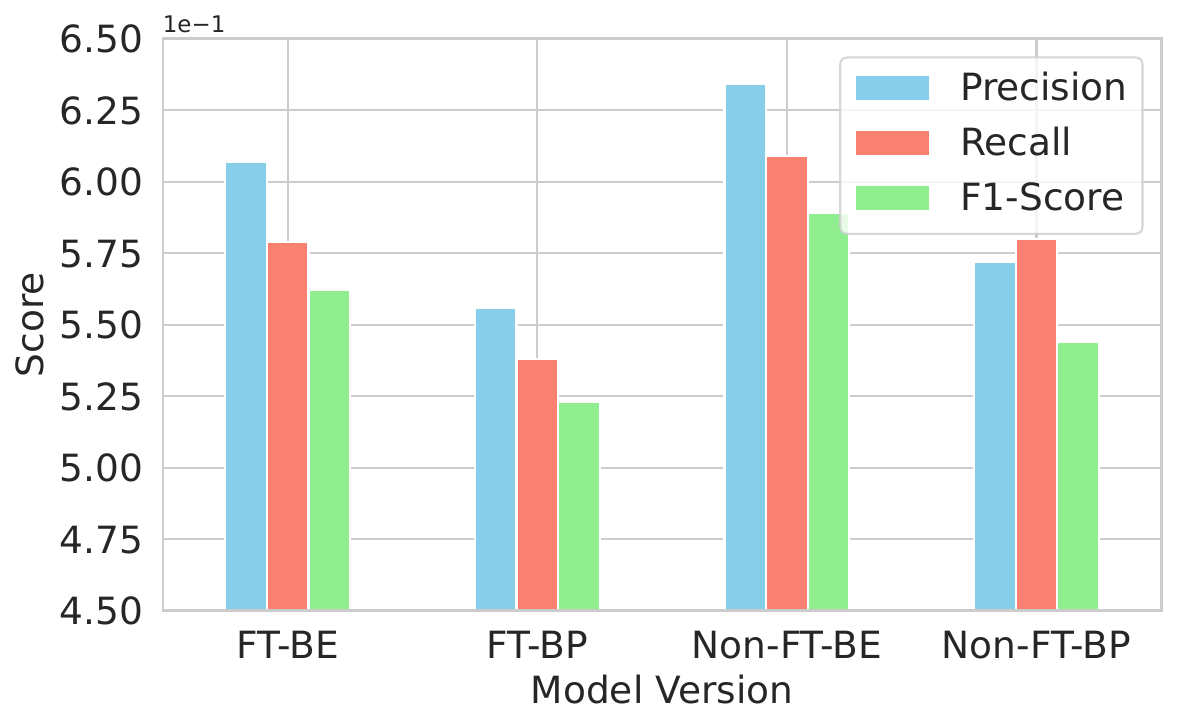}
    \caption{Comparisons of the fine-tuned (FT) and the non-fine-tuned (Non-FT) Llama2 model, each with the best prompt (BP) or the best ensemble (BE)}
    \label{fig:llama_tuning}
\end{figure}

Figure~\ref{fig:llama_tuning} depicts a comparative analysis between fine-tuned (FT) and non-fine-tuned (Non-FT) versions of the Llama2 model, each assessed using both the best prompt and the best ensemble. The depicted results highlight a significant finding, where the Non-FT model leveraging the best ensemble approach outperforms all other configurations in terms of F1 score. In a detailed breakdown of the performance metrics, the Non-FT model with the best ensemble (Non-FT-BE) surpasses the FT model employing the best prompt (FT-BP) by a margin of 11.2\%. It also outperforms the FT model paired with the best ensemble (FT-BE) by 4.6\% and shows a 7.67\% improvement over the Non-FT model that uses the best prompt. For development, the 13 billion parameter Non-FT model was selected. This choice was motivated by the model's computational efficiency, which provides a pragmatic balance between performance and resource expenditure. However, for the final iteration of our results, we capitalized on the superior capacity of the 70 billion parameter Non-FT version. The selection of this model was predicated on its enhanced capability to encode and process complex patterns, thereby optimizing the outcome of our data-cleaning task.

Table~\ref{tab:validation} presents the outcomes of our validation dataset experiments, with a particular focus on measuring precision, recall, and the F1 score. The prompt ensemble method exhibited commendable performance, achieving an average F1-score of 62.53\% across all evaluated relations. A closer examination of the results reveals a notable discrepancy in the predictive success across various relation types. Specifically, the method demonstrated its lowest efficacy on the PersonHasEmployer relation, with an F1-score of 34.97\%, whereas it achieved its highest accuracy on the PersonHasNobelPrize relation, boasting an impressive F1-score of 98.00\%. This variation in performance is likely attributable to the intrinsic differences in the datasets and the distinctive training methodologies applied to the language model (LLM). Relations such as winning a Nobel Prize or establishing a chemical connection represent more distinctive events compared to the commonality of employment relations, which may account for the observed disparity in prediction accuracy. This suggests that the LLM's training may have been more attuned to identifying unique, significant occurrences over more mundane or frequent ones.
\begin{table}[ht]
\centering
\caption{Performance of the prompt ensemble algorithm on multiple relations from the validation set}
\label{tab:validation}
\resizebox{0.9\columnwidth}{!}{%
\begin{tabular}{lccc}
\hline
\textbf{Relation} & \textbf{Precision} & \textbf{Recall} & \textbf{F1-Score} \\ \hline
BandHasMember & 0.6156 & 0.6414 & 0.5920 \\
CityLocatedAtRiver & 0.6900 & 0.6048 & 0.6099 \\
CompanyHasParentOrganisation & 0.8700 & 0.6150 & 0.5867 \\
CompoundHasParts & 0.9780 & 0.9755 & 0.9747 \\
CountryBordersCountry & 0.8248 & 0.8402 & 0.8038 \\
CountryHasOfficialLanguage & 0.8949 & 0.8346 & 0.8413 \\
CountryHasStates & 0.5770 & 0.7115 & 0.6214 \\
FootballerPlaysPosition & 0.6050 & 0.7433 & 0.6413 \\
PersonCauseOfDeath & 0.7000 & 0.7433 & 0.6950 \\
PersonHasEmployer  & 0.4163 & 0.3777 & 0.3497 \\
PersonHasNoblePrize & 0.9900 & 0.9900 & 0.9800 \\
PersonHasSpouse & 0.6800 & 0.6600 & 0.6633 \\
PersonSpeaksLanguage & 0.9008 & 0.7702 & 0.7856 \\
RiverBasinsCountry & 0.8123 & 0.8529 & 0.7803 \\ \hline
\end{tabular}
}
\end{table}


\subsubsection{Error Detection Results} 
\begin{figure*}[htbp] 
	\centering
    \subfloat[IoT dataset]{\label{fig:accuracy_iot}\includegraphics[width=0.33\textwidth]{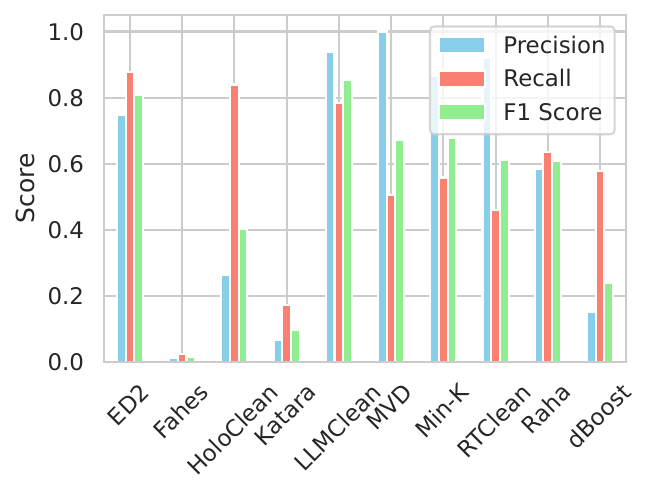}}	%
	\subfloat[Hospital dataset]{\label{fig:accuracy_hospital}\includegraphics[width=0.33\textwidth]{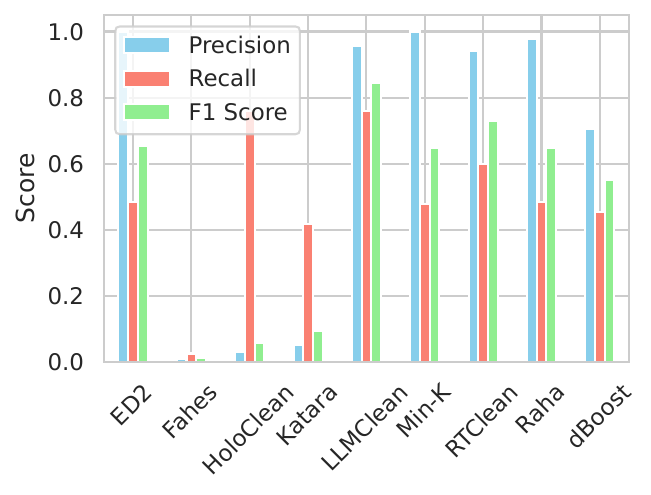}}
	\subfloat[Context dataset]{\label{fig:accuracy_context}\includegraphics[width=0.33\textwidth]{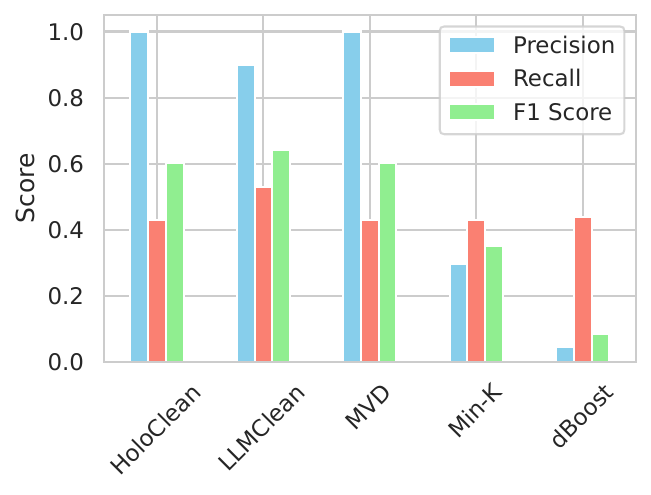}}
	\caption{Accuracy of error detection comparing \PaperAcronym to the baselines}
 	\label{fig:accuracy} 
 \end{figure*}

Figure~\ref{fig:accuracy} depicts the accuracy of \PaperAcronym and the compared baseline tools--in terms of the detection precision, recall, and F-Score--while detecting errors in three real-world datasets. For the IoT datasets, Figure~\ref{fig:accuracy_iot} shows that \PaperAcronym demonstrates superior performance in terms of the F1-score when compared to various baseline tools. It notably surpasses HoloClean, ED2, Pandas' Missing Value Detector (MVD), and Raha with substantial margins of improvement--53\%, 5.4\%, 21.3\%, and 28.7\% respectively. Further underlining its efficiency, \PaperAcronym identified 868 cells as containing errors, significantly lower than the 1222, 1131, and 3316 instances flagged by ED2, Raha, and HoloClean, respectively. These figures not only highlight the precision of \PaperAcronym but also its effectiveness in accurately detecting erroneous data within a dataset.

\begin{figure*}[htbp] 
	\centering
    \subfloat[IoT dataset]{\label{fig:runtime_iot}\includegraphics[width=0.33\textwidth]{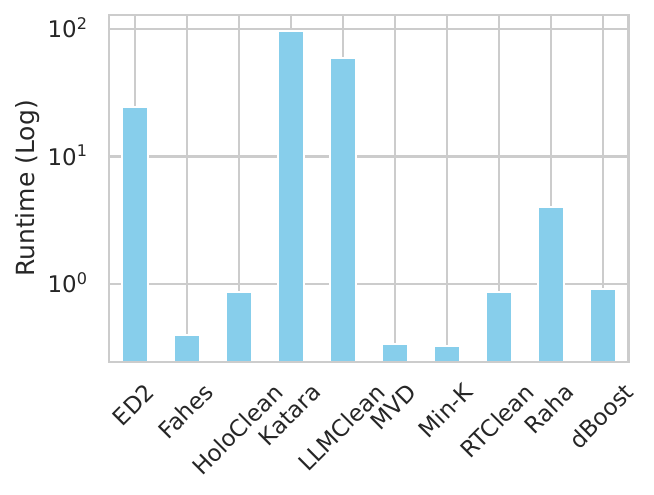}}	%
	\subfloat[Hospital dataset]{\label{fig:runtime_hospital}\includegraphics[width=0.33\textwidth]{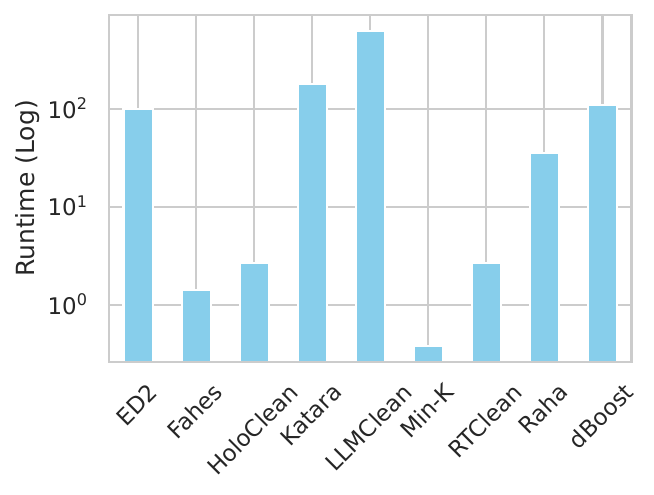}}
	\subfloat[Context dataset]{\label{fig:runtime_context}\includegraphics[width=0.33\textwidth]{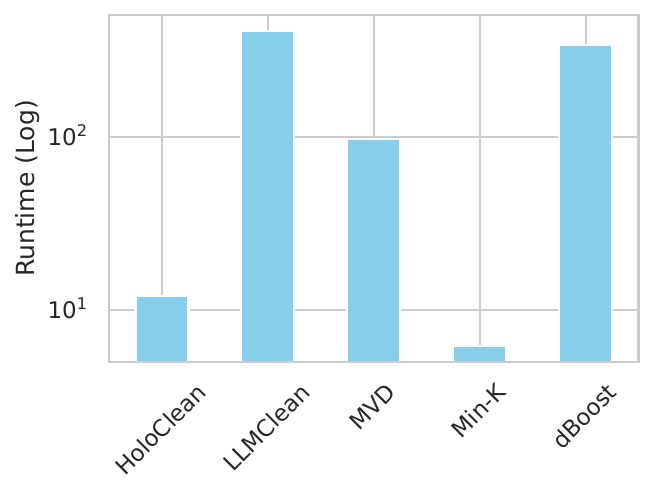}}
	\caption{Runtime of error detection comparing \PaperAcronym to the baselines}
 	\label{fig:runtime} 
 \end{figure*}

\begin{figure*}[htbp] 
	\centering
    \subfloat[Numerical IoT]{\label{fig:num_iot}\includegraphics[width=0.25\textwidth]{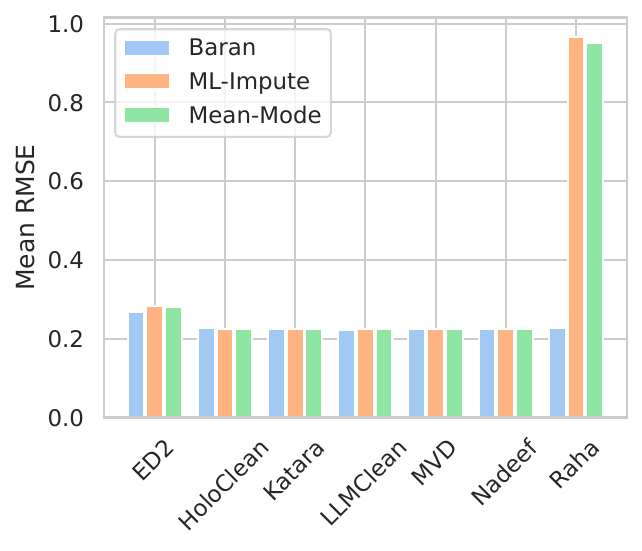}}	%
	\subfloat[Categorical IoT]{\label{fig:cat_iot}\includegraphics[width=0.25\textwidth]{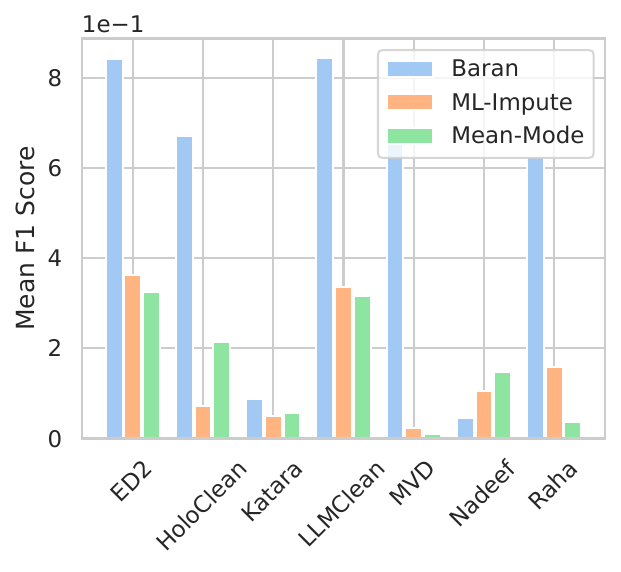}}
	\subfloat[Numerical Hospital]{\label{fig:num_hospital}\includegraphics[width=0.25\textwidth]{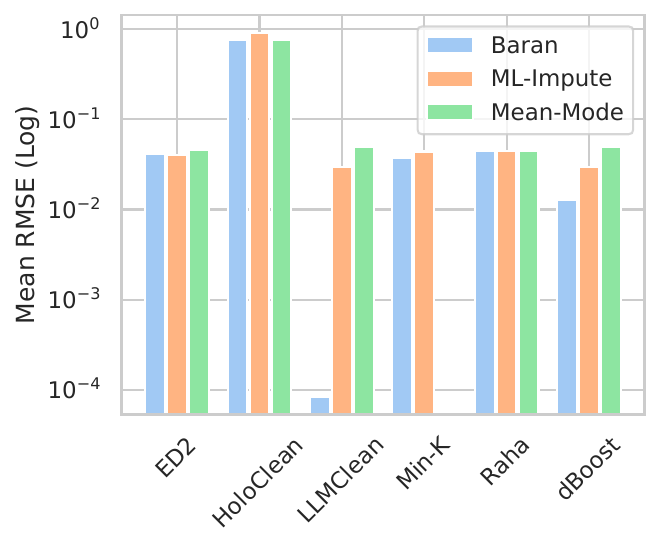}}
    \subfloat[Categorical Hospital]{\label{fig:cat_hospital}\includegraphics[width=0.25\textwidth]{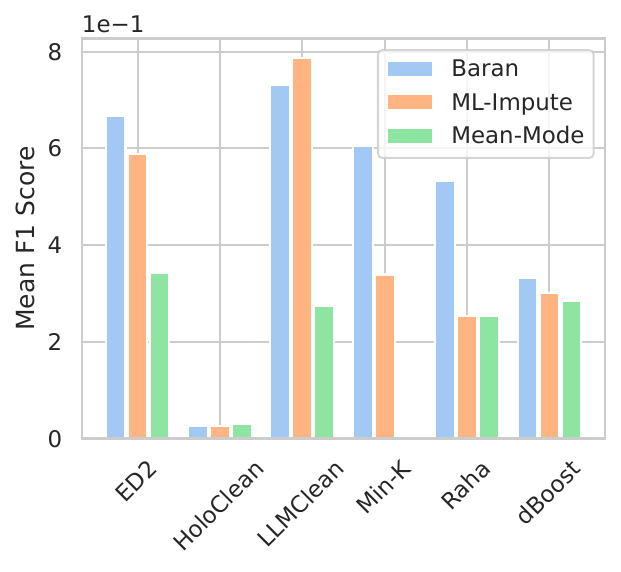}}
	\caption{Accuracy of error repair comparing \PaperAcronym to the baselines}
 	\label{fig:repair} 
 \end{figure*}

For the Hospital dataset, \PaperAcronym{} demonstrates superior performance over conventional baseline tools when it comes to the detection F1-score. To illustrate, \PaperAcronym{} surpasses ED2, RAHA, and dBoost by substantial margins of 22.7\%, 23.2\%, and 34.7\%, respectively. A closer look at the results reveals that \PaperAcronym{} identified 404 cells as erroneous, which represents a significant increase in detection over the 253, 247, and 328 cells flagged by RAHA, ED2, and dBoost, respectively. Interestingly, even when operating under the same number of FD rules as HoloClean, \PaperAcronym{} vastly overshadows its performance. HoloClean detected 13,044 cells, a number which, due to its magnitude, severely diminished its F1-score to less than 1\%, highlighting the precision and efficiency of \PaperAcronym{}. For the CONTEXT dataset, ML-based tools, such as RAHA and ED2, encountered operational challenges, specifically, they were unable to complete execution due to the dataset's extensive size. Yet again, \PaperAcronym{} stands out, outstripping all baseline tools with an F1-score that is 6.2\% higher in comparison to both MVD and HoloClean. This consistent outperformance across different datasets underscores the robustness of \PaperAcronym{}.

Figure~\ref{fig:runtime} presents the runtime analysis of \PaperAcronym in comparison with baseline tools. From the figure, it is evident that \PaperAcronym's runtime is marginally higher than that of the most competitive baselines. The reason for this could be attributed to \PaperAcronym's thorough approach, which involves checking and validating numerous combinations of column pairs. Additionally, the runtime for \PaperAcronym is largely influenced by the quantity of generated OFD rules. For instance, in the case of the IoT dataset, \PaperAcronym takes approximately 0.98 minutes to apply two OFD rules. However, when applying 21 OFD rules to the Hospital dataset, the runtime extends to approximately 10.56 minutes. A similar pattern is observed with the CONTEXT dataset, where \PaperAcronym spends 6.8 minutes to thoroughly examine the data. This contrasts with the 5.6 minutes taken by dBoost and a swift 1.6 minutes by the MVD detector. These runtime variances highlight the complexity and depth of analysis conducted by \PaperAcronym, particularly concerning the number of OFD rules it enforces.

\subsubsection{Error Repair Results} While \PaperAcronym is primarily deployed for error detection within datasets, assessing the effectiveness of its detection in conjunction with subsequent repair processes is essential. This section provides an analysis of the combined performance of error detection and repair, utilizing \PaperAcronym and various baseline tools alongside three SOTA repair tools: Baran, standard statistical imputation, and ML-based imputation. For statistical imputation, we apply mean-value imputation for numerical attributes and mode-value imputation for categorical ones. The ML-based imputation employs a K-Nearest Neighbors (KNN) regressor for numerical attributes and MissForest for categorical attributes. An examination of the evaluation results for the IoT and Hospital datasets, presented in Figure~\ref{fig:repair}, reveals noteworthy outcomes\footnote{The CONTEXT dataset analysis is omitted here for conciseness.}. Specifically, Figure~\ref{fig:num_iot} indicates that \PaperAcronym consistently achieves the lowest RMSE of 0.22 across numerical attributes, independent of the repair mechanism employed. This performance is comparable to the results of other tools, such as HoloClean, MVD, and Nadeef.

Focusing on the categorical attributes of the IoT dataset, as shown in Figure~\ref{fig:cat_iot}, the combination of \PaperAcronym and Baran attains the highest F1-score at 85\%. This marginally surpasses the 84\% F1-score observed with ED2 and Baran, and significantly outperforms the 67\% F1-score seen with MVD and Baran. For the numerical attributes of the Hospital dataset, as depicted in Figure~\ref{fig:num_hospital}, the pairing of \PaperAcronym with Baran again yields the most favorable RMSE value (8.44E-05). It is important to note that the vertical axis of this figure is in a logarithmic scale to appropriately represent the notably small values achieved by this combination. Lastly, Figure~\ref{fig:cat_hospital} demonstrates that for the Hospital dataset's categorical attributes, \PaperAcronym used with ML-based imputation (ML-Impute) secures the highest F1-score of 78.7\%. This result outdoes the 73\% F1-score obtained with \PaperAcronym and Baran, as well as the 66.7\% achieved by combining ED2 with Baran. To sum up, these findings underscore the superior error detection and repair capabilities of \PaperAcronym, particularly when combined with advanced repair methodologies across both numerical and categorical data domains.

\subsection{Discussion}
\textbf{Context Generation:} The evaluation revealed that the process of automatically generating context models for tabular datasets is highly effective. In IoT datasets, the data is often generated by sensors and devices with known and fixed schemas, which means that the context is well-defined and the types of possible dependencies (like Sensor Capability Dependencies and Device-Link Dependencies) can be anticipated and modeled accordingly. This makes the automatic generation of context models more straightforward because LLM models can be trained to recognize these regular patterns and dependencies with high accuracy. For non-IoT datasets, it can become challenging for LLM models to generate context models since they can come from a multitude of sources with less structured and more heterogeneous contexts. However, the evaluation of the Hospital dataset (Figures~\ref{fig:accuracy_hospital},\ref{fig:num_hospital}, and \ref{fig:cat_hospital}) indicates that \PaperAcronym sustains its effectiveness in both error detection and repair tasks within this category of datasets.

\textbf{Contextual Changes:} To evaluate \PaperAcronym's adaptability to contextual changes, we compared F1 scores before and after a simulated context change in the IoT dataset, where \quotes{Room2} is renamed to \quotes{Room3} and the associated device and sensor labels are updated accordingly. We found that \PaperAcronym maintains similar performance in error detection and repair despite the contextual alterations. Accordingly, we can conclude that if the context changes are limited to a single sensor rather than a series of sensors, or if they are not distributed across a more extended period, this could influence the adaptability and the necessity for modifications to the context model. Consequently, a restricted scope of alterations or a lack of frequent, diverse changes over time leads to a decreased need for model adaptation.


\textbf{OFD Dependencies:} In the absence of these Capability Dependencies, \PaperAcronym may overlook specific interrelations or dependencies among data instances, which can lead to a decline in the accuracy of identifying errors. The generation process of such dependencies is markedly more effective when the sensor's name is explicitly mentioned in the dataset, enabling the system to harness additional information, such as technical specifications, from online resources. The explicit mention of the sensor name is a key enabler for accurately aligning sensor capabilities with their respective data instances. However, the approach's reliance on the availability of technical specifications online casts light on its dependency on external data sources, with the quality and precision of the automatically generated sensor capabilities being directly proportional to the richness and accuracy of information available on the Internet. Remarkably, our findings suggest that the lack of Temporal and Monitoring Dependencies within the context model seemingly does not compromise the results, a phenomenon that could be related to the unique characteristics of the dataset at hand.

\textbf{Prompt Engineering}: The evaluation of our prompt engineering approach for the LM-KBC dataset shows that larger parameter sizes, more few-shot examples, and the application of Prompt Ensembling lead to better results. Despite its advantages, the LLM approach also has some drawbacks. In particular, it is not very stable, which means that the results can vary between different runs or in different scenarios. This instability could be due to various factors, such as the model's sensitivity to small changes in the input data or dependence on certain parameter settings. To improve the stability of the results, the method of prompt ensembling in combination with sampling is used. By applying sampling, random variations in the training data can be simulated, which can improve model generalization. The combination of prompt ensembling and sampling thus represents an approach to increase the stability of the LLM and achieve more consistent results, particularly in situations where the LLM model shows weaknesses in terms of stability.

\section{Related Work}\label{sec:related_work}
%
In this section, we provide an overview of the literature in the realm of automated data cleaning and the generation of OFD rules. We explore recent advancements together with highlighting the unique contributions that \PaperAcronym{} offers data quality and integrity.

\subsection{Context Model Generation}
The field of context model generation is characterized by a diverse array of approaches that can be broadly categorized into two groups: ML-agnostic and ML-based solutions.

\textbf{ML-agnostic Solutions} are those that do not utilize ML algorithms. Instead, they rely on other computational methods such as rule mining or heuristic algorithms to interpret and structure data. For instance, Sateli et al.~\cite{sateli2015automatic} proposed a method for constructing knowledge bases by leveraging the semantic content of scholarly publications. This discipline is dedicated to enhancing the accessibility of scientific literature, ensuring that it is interpretable not only by human readers but also by machines. The particular challenge addressed by Sateli et al. is the extraction of relevant information from academic journal articles. The information extracted from these articles is systematically organized into an RDF graph. Such a method does not rely on machine learning techniques but on rule mining. The extraction process is methodically broken down into three primary stages: syntactic processing, semantic processing, and the subsequent exportation of annotations. In the initial syntactic processing phase, the text document is transformed into discrete elements known as tokens. These tokens then undergo a comparative analysis during the semantic processing stage against a pre-compiled list of tokens. This list is curated, containing tokens that are specifically designed to align with the information being sought. Subsequently, through the employment of rule transducers, the target information is extracted from the sentences. 

Similarly, Kertkeidkachorn et al.~\cite{kertkeidkachorn2018automatic} introduce an approach for the automatic generation of knowledge graphs from natural language texts. A key component of their approach is the \quotes{Predicate Mapping} process, which involves aligning predicates found within the text to their corresponding entities within a knowledge graph. This method is a hybrid one, combining rule-based mechanisms with similarity-based techniques. This fusion is designed to enhance the enrichment of triples—which consist of subject, predicate, and object—from the text to be integrated into an extant knowledge graph. By doing so, the method aims to capture knowledge with higher precision and effectively incorporate it into the knowledge graph. By effectively mapping natural language predicates to their knowledge graph counterparts, the process ensures a more seamless and coherent integration of information, enriching the existing graph with new and relevant data. 

While ML-agnostic solutions can be useful in certain scenarios, they have several limitations. Such tools are usually rule-based and lack the flexibility of ML models that can learn and adapt from data over time. This makes them less capable of handling new, unforeseen scenarios that fall outside their predefined rules. Creating and maintaining such rules requires expert knowledge and can be time-consuming. This also means that the quality of the context model is highly dependent on the expertise of the rule creators. Moreover, if the rules are incorrectly defined, errors can propagate throughout the system, leading to inaccurate context models. Finally, such solutions may struggle with complex data relationships that are not easily defined by straightforward rules.

\textbf{ML-based solutions}, On the other hand, employ ML techniques to derive context models from raw data. For example, Carta et al.~\cite{carta2023iterative} presented a new iterative zero-shot technique that stands out by not relying on external knowledge bases, instead it harnesses the intrinsic capabilities of LLM models through a series of refined prompts. Each iteration builds upon the last, negating the need for example-driven guidance and streamlining the path to accurate component extraction. Along a similar line, Trajanoska et al.~\cite{trajanoska2023enhancing} delve into the capabilities of foundational models like GPT and specialized models such as Rebel for generating Knowledge Graphs from unstructured sustainability texts. They assess the efficacy of these models in two key areas: the extraction of relationships between concepts, and the integration of new concepts into an existing ontology. 

The comparative study unfolds across three experimental setups to provide a comprehensive analysis. First, Rebel is tasked with relation extraction and DBpedia with entity linking, establishing a benchmark for specialized model performance. Second, ChatGPT is employed for relation extraction, maintaining consistent entity linking with DBpedia to ensure a direct comparison. This allows for an evaluation of how a conversational AI fares in a specialized task. Lastly, GPT is challenged to build an entire ontology, starting from basic sustainability concepts, where ChatGPT further enriches this ontology. The study determined that the Knowledge Graph developed through the second setup, i.e., the ChatGPT approach, surpasses the quality of the other two Knowledge Bases. 

\subsection{Automated Data Cleaning}
This section provides an overview of the advancements in automating the cleansing of tabular datasets. To facilitate a structured discussion, the solutions are clustered into two primary categories: rule-based cleaning tools and ML-based cleaning tools.

\textbf{Rule-Based Cleaning Tools} rely on predefined rules and logic to identify and rectify inconsistencies, errors, or anomalies in tabular data. For instance, Zheng et al.~\cite{zheng2021discovery} explored the use of OFD dependencies for enhancing data-cleaning processes. They recognize that real-world data often contain complex, domain-specific relationships that simple syntactic rules cannot capture, such as the presence of synonyms and other semantic connections. To leverage these semantic relationships, the authors turn to ontologies, which offer structured models of dataset semantics. They introduced an approach where OFDs are defined in terms of synonym relationships derived from an ontology. Zheng et al. present an innovative algorithm designed to discover these OFDs by mining the ontology for synonymous terms. The subsequent data cleaning algorithm, OFDClean, utilizes the discovered OFDs to determine the most accurate interpretation for a group of tuples considered semantically equivalent. 

Zheng et al. note that the effectiveness of OFDs in identifying errors varies across different types of datasets. In the context of IoT environments, where data is predominantly numerical, semantic synonyms are less relevant for error detection. Since numerical values do not typically have synonyms or semantic equivalence classes, the application of OFDs is somewhat limited. To address the gap in IoT data cleaning, \PaperAcronym{} incorporates contextual information from ontologies into predominantly numerical datasets. These proposed concepts aim to extend the utility of OFDs, allowing them to play a more significant role in identifying and correcting errors in various types of datasets, not just those that are text-based. 

In \cite{rtclean23}, we introduced RTClean, a novel data-cleaning method that takes into account the context in which data is collected and adapts to the dynamic nature of deployment environments. By utilizing a \textit{live} context model tailored to the application's needs, RTClean captures essential OFD dependencies to inform its cleaning process. What sets RTClean apart is its capacity to incorporate real-time environmental changes through continuous integration of live data from monitoring systems and sensors. This ensures that the context model remains current, reflecting the ever-changing landscape of available devices and sensors, and thereby maintaining the relevance and accuracy of data cleaning efforts.
Aside from OFDs, Rekatsinas et al. \cite{holoclean17} introduced HoloClean, a system that integrates integrity constraints, external datasets, and statistical methods to identify and rectify errors in structured data. It constructs a probabilistic model that encapsulates the uncertainty within the dataset's tuples. This model transforms signals into features that define the data's probabilistic characteristics. For error correction, HoloClean employs statistical learning and inference, leveraging the probabilistic model to detect and repair inaccuracies. 
 
\textbf{ML-Based Cleaning Tools}, on the other hand, employ ML models to learn from the data itself \cite{rein23,autocure2023,saged24,raha19,holodetect19}. For instance, SAGED is a meta-learning tool crafted for detecting errors in tabular data. This tool leverages meta-learning techniques, which utilize insights from previously trained models on related tasks to improve learning in new tasks or domains, especially when labeled data is scarce. SAGED's design draws on the knowledge of pre-cleaned historical datasets, allowing it to perform error detection swiftly and accurately. Similarly, RAHA \cite{raha19} is a configuration-free error detection tool that exploits semi-supervised learning to train a set of detection classifiers. Recently, several efforts have been exerted to leverage LLM models in data engineering tasks. For instance, Narayan et al. \cite{narayan2022can} successfully applied a prompting strategy that incorporates examples, i.e., a few-shot approach, to improve the performance of LLMs, surpassing traditional ML-based tools. Complementarily, Vos et al. \cite{vos2022towards} explored prefix-tuning as a lightweight alternative to the more resource-heavy fine-tuning for adapting LLMs to data-wrangling tasks. Furthering this innovation, RetClean \cite{ahmad2023retclean} was developed to refine ChatGPT's outputs using data from a specified data lake. While LLMs show promise for data management, their practical application is nascent and fraught with challenges, including the need for domain-specific adaptations, data privacy issues, and substantial computational demands.  

%
\section{Conclusion \& Future Work}\label{sec:conclusion}
%
This paper introduces a novel method leveraging LLM models to autonomously extract context models from datasets, requiring no supplementary dataset information. The necessity for automated context model generation arises from the labor-intensive nature of manual creation. These context models are crucial for data cleaning, and preparing the data for subsequent AI applications, and OFD dependencies play a supportive role in this process. Our approach, \PaperAcronym{}, streamlines the creation of context models for integration into data-cleaning workflows. We categorized two dataset types, emphasizing IoT datasets replete with sensor data, while the second deals with non-IoT data in relational databases. Our evaluation shows that the context models generated by \PaperAcronym{} are effective in cleansing both dataset types. Several avenues for future research hold considerable potential. Enhancing table-to-knowledge graph conversions by incorporating knowledge graph embeddings could significantly refine the accuracy and semantic richness of these transformations. Exploring embeddings for non-IoT relational data presents another interesting prospect; these embeddings could more accurately represent semantic interconnections between entities, thereby elevating model precision.



\begin{acks}
This research was funded by the German Federal Ministry of Education and Research (BMBF) through grants 01IS17051 (Software Campus program), 02L19C155, and 01IS21021A (ITEA project number 20219).
\end{acks}

\bibliographystyle{ACM-Reference-Format}
\bibliography{main.bib}

\end{document}